\documentclass{article}

\usepackage{arxiv}

\usepackage[utf8]{inputenc} 
\usepackage[T1]{fontenc}    
\usepackage{hyperref}       
\usepackage{url}            
\usepackage{booktabs}       
\usepackage{amsfonts}       
\usepackage{nicefrac}       
\usepackage{microtype}      

\usepackage{graphicx}
\usepackage{bm}
\usepackage{amsmath}
\usepackage{amssymb}
\usepackage{amsfonts}

\usepackage{algorithm}
\usepackage{algorithmic}

\usepackage{breqn}

\title{Symbolic-Numeric Integration of Univariate Expressions based on Sparse Regression}

\author{
    Shahriar Iravanian\thanks{shahriar.iravanian@emoryhealthcare.org}\\
    Emory University
    \And
    Carl Julius Martensen \\
    Otto-von-Guericke University \\
    \And
    Alessandro Cheli \\
    University of Pisa \\
    \And
    Shashi Gowda \\
    Massachusetts Institute of Technology \\
    \And
    Anand Jain \\
    Julia Computing \\
    \And
    Yingbo Ma \\
    Julia Computing \\
    \And
    Chris Rackauckas \\
    Massachusetts Institute of Technology 
}

\begin{document}
\maketitle
\begin{abstract}
Most computer algebra systems (CAS) support symbolic integration as core functionality. The majority of the integration packages use a combination of heuristic algebraic and rule-based (integration table) methods. In this paper, we present a hybrid (symbolic-numeric) methodology to calculate the indefinite integrals of univariate expressions. The primary motivation for this work is to add symbolic integration functionality to a modern CAS (the symbolic manipulation packages of SciML, the Scientific Machine Learning ecosystem of the Julia programming language), which is mainly designed toward numerical and machine learning applications and has a different set of features than traditional CAS. The symbolic part of our method is based on the combination of candidate terms generation (borrowed from the Homotopy operators theory) with rule-based expression transformations provided by the underlying CAS. The numeric part is based on sparse-regression, a component of Sparse Identification of Nonlinear Dynamics (SINDy) technique. We show that this system can solve a large variety of common integration problems using only a few dozen basic integration rules.
\end{abstract}


\section{Introduction}

Symbolic integration is a core competency of most Computer Algebra Systems (CAS) and has numerous applications~\cite{Bronstein2005, Lamagna2018, Gerhard2011}. Since the advent of symbolic integration software in the 1960s~\cite{R.1963, Moses}, the majority of the integration packages fall on a spectrum between algebraic methods - based on a combination of heuristics such as derivative-divide and integration-by-part methods to the Risch algorithm - and rule-based methods, incorporating thousands of specific integration rules (e.g., RUBI)\cite{Risch1969, Meurer2017, Rich2018}. In this paper, we develop a symbolic-numeric (hybrid) integration method, which is closer in spirit to the heuristic methods but takes a detour in numerical computation to simplify the intermediate steps.

\textbf{JuliaSymbolics} is a subset organization of \textbf{SciML} (an open source organization maintaining a collection of hundreds of scientific computing packages for the Julia programming language). Historically, \textbf{SciML} began as a group of ordinary differential equation (ODE) solvers~\cite{rackauckas2017differentialequations}. Symbolic computation was added as an Embedded Domain Specific Language (eDSL) to ease the definition of ODE systems and to aid with automatic calculation of Jacobian and Hessian of such systems. With the expansion of the \textbf{SciML} ecosystem, the purely symbolic routines were decoupled from the ODE solvers and were collected under \textbf{JuliaSymbolics}~\cite{gowda2021high}. Considering its history and origin, \textbf{JuliaSymbolics} is geared toward symbolic differentiation and numerical integration but lacks direct symbolic integration capabilities. Furthermore, \textbf{SciML} has grown to cover recent advances in scientific machine learning. As part of our method, we utilize one of the these new packages (\textbf{DataDrivenDiffEq.jl}) that implements data-driven differential equation structural estimation and identification~\cite{datadrivendiffeq}.

\textbf{SymbolicNumericIntegration.jl} is an attempt to build a symbolic integration package by combining the symbolic manipulation and automatic differentiation capabilities of \textbf{JuliaSymbolics} with numerical routines such as sparse regression provided by \textbf{DataDrivenDiffEq.jl}.

\section{The Overview of the Method}

This section informally introduces the symbolic-numeric integration algorithm by a simple example. We want to find $\int f(x)\,dx$, where $f : \mathbb{C} \rightarrow \mathbb{C} = x \sin x$. We can integrate $f$ using the \emph{method of indeterminate coefficients}. The main idea is to write the solution, i.e., $S = \int x \sin x\,dx$, as a sum of multiple possible terms with unknown coefficients,
\begin{equation}
  S = \sum_i q_i \theta_i(x)
  \,,
\end{equation}
where $q_i \in \mathbb{C}$, for $i = 1\cdots|T|$, are the unknown coefficients and $T = \{\theta_i(x)\}$, where $\theta_i: \mathbb{C} \rightarrow \mathbb{C}$, is the set of \emph{reasonable candidate} terms (\emph{ansatz}). For our example, a reasonable set of terms is $T = \{x, \sin x, \cos x, x\sin x, x\cos x\}$. Of course, we need a better method to find $T$ than saying it should be a reasonable set! We will discuss this problem is details later (Section 3.2), but for now assume that an oracle provides $T$. We have
\begin{equation}
  S = q_1 x + q_2 \sin x  + q_3 \cos x + q_4 x \sin x + q_5 x \cos x
  \,.
\end{equation}
Differentiating with respect to $x$,
\begin{equation}
  S' = q_1 + (q_4 - q_3) \sin x + (q_2 + q_5) \cos x - q_5 x \sin x + q_4 x \cos x
  \,.
\end{equation}
By definition, $S = \int f\,dx$; therefore, $S' = f = x \sin x$. We obtain the following linear system,
\begin{equation}
  \begin{array}{ll}
    q_1 = 0 \\
    q_4 - q_3 = 0 \\
    q_2 + q_5 = 0 \\
    -q_5 = 1 \\
    q_4 = 0  
  \end{array}  
\end{equation}
The solution to the linear the system is $q_5 = -1$, $q_2 = 1$, and $q_1 = q_3 = q_4 = 0$. Therefore,
\begin{equation}
  S = \int x \sin x\, dx = \cos x - x \cos x 
  \,.
\end{equation}
As it should. 

Note that the preceding calculations were essentially all symbolic. However, numerical computation becomes necessary to avoid relying on \textbf{JuliaSymbolics} to convert expressions into unique \emph{canonical} forms. Identities like $\sin^2 x + \cos^2 x = 1$ may be correctly applied in this case, but in general, according to Richardson's theorem, the problem of finding canonical forms of transcendental expressions is undecided~\cite{Richardson1969}. Another reason for using numerical computation is that the list of candidates may not be (and usually is not) linearly independent. Finding a linearly-independent subset of a set of expressions is facilitated using numerical methods (Section 4.3). 

\section{Symbolic Computation}
\subsection{The Main Integration Algorithm}

The main symbolic-numeric integration algorithm is summarized in Algorithm~\ref{alg:main}. Let $x$ be the independent variable. For this paper, we assume that the input function to be integrated, $f : \mathbb{C} \rightarrow \mathbb{C}$, is a univariate expression of $x$. Additionally, we assume that $f$ is well defined in a closed subset of the complex plane with only isolated poles.

The candidates generation algorithm (Section 3.2) produces a list of generator expressions, $G_0, G_1, G_2,\cdots$ (by applying Equations~\ref{eq:prod_u}, \ref{eq:B0} and \ref{eq:Bi}), where $G_i = \sum_{k} a_k \theta_k(x)$, $a_k \in \mathbb{C}$, $\theta_k: \mathbb{C} \rightarrow \mathbb{C}$, and $\theta_k$ does not have a constant leading coefficient. Each $G_i$ is converted to a set of candidates, $T_i = \{\theta_k\}$. According to Eq.~\ref{eq:Bi}, $T_i \subset T_{i+1}$.  We can ignore the constant leading coefficients because the final coefficients are calculated by the numerical part of the algorithm. Therefore, we can express $\theta_k$s as equivalence classes of expressions where $f(x) \sim g(x)$ iff $f/g$ is a non-zero constant. Using this convention, we may write $(\sin x)' = \cos x$ and $(\cos x)' = \sin x$. 

\begin{algorithm}
\caption{Symbolic-Numeric Integration}
\begin{algorithmic}[1]
\REQUIRE A univariate expression $f(x)$ with constant coefficients in $\mathbb{C}$
\REQUIRE $L$, a user-defined parameter that determines the number of generators to try
\ENSURE A symbolic expression $y$, the antiderivative of $f$, such that $y' = f$
\FOR{$k = 0\dots L-1$} 
    \STATE $G_k \gets$ the $k$th candidate generator
    \COMMENT {Eqs.~\ref{eq:prod_u}, \ref{eq:B0}, and \ref{eq:Bi}}
    \STATE $T_k \gets$ the candidate terms of $G_k$
    \STATE $y \gets$ the anti-derivative of $f$ using $T_k$ \COMMENT{Algorithm 2}
    \IF{$y'$ is equal to $f$}
        \RETURN $y$
    \ENDIF
\ENDFOR
\RETURN \textbf{no answer}
\end{algorithmic}
\label{alg:main}
\end{algorithm}

For each $G_i$ and the corresponding $T_i$, Algorithm~\ref{alg:main} calls Algorithm~\ref{alg:num} to find the anti-derivative of the input. If there is an acceptable solution, it returns; otherwise, Algorithm~\ref{alg:main} fetches the next generator, $G_{i+1}$, and tests the resulting candidate set, $T_{i+1}$. If no solution is found after trying $G_{L-1}$, it returns with a failure message. $L$ is a parameter specified by the user. 

\begin{algorithm}
\caption{The Numerical part of Symbolic-Numeric Integration}
\begin{algorithmic}[1]
\REQUIRE A univariate expression $f(x)$ with constant coefficients in $\mathbb{C}$
\REQUIRE A set of candidate terms $T = \{\theta_1,\theta_2,\cdots,\theta_n\}$
\ENSURE A symbolic expression $y$, the antiderivative of $f$, such that $y' = f$
\FOR{$i \gets 1\dots n$}
    \STATE $x_i \gets$ a random complex numbers inside the open disk $\mathbb{D}_d$
    \STATE move $x_i$ toward the poles of $f$
    \STATE $\mathbf{b}_{i} \gets f(x_i)$ \COMMENT{$\mathbf{b}$ is a vector of $n$ elements}    
    \FOR{$j \gets 1\dots n$}
        \STATE $\mathbf{A}_{ij} \gets \theta'_j(x_i)$ \COMMENT{$\mathbf{A}$ is an $n$-by-$n$ matrix and $\theta'_j$ is the derivative of $\theta_j$ with respect to $x$}
    \ENDFOR
\ENDFOR
\STATE Find $R \subset \{1,2,\dots,n\}$, the set of linearly-dependent rows of $\mathbf{A}$
\FOR{$i \in R$}
    \STATE remove the $i$th row and column from $\mathbf{A}$
    \STATE remove the $i$th row from $\mathbf{b}$
\ENDFOR
\STATE solve $\min_q \lVert \mathbf{A} \mathbf{q} - \mathbf{b} \rVert_2^2 + \lambda \lVert \mathbf{q} \rVert_2$ for $\mathbf{q}$ (sparse regression)
\STATE $y \gets \sum_j q_j \theta_j$ 
\COMMENT{$\mathbf{A}$ uses $\theta'$, but $y$ is based on $\theta$}
\RETURN y
\end{algorithmic}
\label{alg:num}
\end{algorithm}

Algorithm~\ref{alg:num} generates $n$ test points $x_i$ in complex plane, where $n$ is the number of the candidates, to create an $n$-by-$n$ matrix $\mathbf{A}$ and an $n$-element vector $\mathbf{b}$ filled, respectively, with the values of the derivatives of the candidate terms and the input function at the test points. 

As discussed above, a potential complication at this stage is that the rows of $\mathbf{A}$ may be linearly-dependent. We remedy this problem by finding the linearly-dependent rows and removing them from $\mathbf{A}$ and $\mathbf{b}$ (Section 4.3).

Using full-rank $\mathbf{A}$ and $\mathbf{b}$, we find $\mathbf{q}$ such that $\mathbf{A}\mathbf{q} = \mathbf{b}$. As is discussed in Section 4.4, we cast the problem as an optimization task and use sparse regression to calculate $\mathbf{q}$. Finally, we put everything together and generate $y = \sum_j q_j \theta_j$, the candidate anti-derivative of the input. 

\subsection{Candidates Generation}

Candidate generation is the core of the Symbolic-Numeric integration algorithm. First, we describe an algorithm for a subset of expressions amenable to simple treatment and then expand to the general algorithm.

One key observation is that the form of the anti-derivative of some functions is similar to their derivative forms. For example, $\int \cos x\,dx = \sin x = -(\cos x)'$. The underlying reason is that these functions can be defined in term of the polynomials of the exponential function and its inverse, i.e., $f \in \mathbb{C}[e^x,e^{-x}]$, $(e^x)' = e^x$, and $(e^{-x})' = -e^{-x}$; therefore $\mathbb{C}[e^x,e^{-x}]$ is closed under differentiation and integration. The functions with this property include the non-negative integer powers of $\sin x$, $\cos x$, $\sinh x$, $\cosh x$, and $e^x$. Let's define
\begin{equation}
    f(x) = (e^x)^{p_1} (\sin x)^{p_2} (\cos x)^{p_3} (\sinh x)^{p_4} (\cosh x)^{p_5}
    \,,
    \label{eq:def_exact}
\end{equation}
where $p_k \in \mathbb{Z}^+$. For these class of functions, we can obtain the candidate terms with repetitive differentiation. For example, let $f(x) = e^x\sin x$. We start with $T = \emptyset$ ($T$ is the set of candidates), differentiate $f$, and add the resulting terms to $T$ (ignoring the constant coefficients):
\begin{equation}
    f' = e^x\cos x + e^x\sin x \Rightarrow T = \{e^x\cos x, e^x\sin x\}
    \,.
\end{equation}
Next, we need to differentiate each term in $T$ that was not previously processed. The first term, $e^x\cos x$, gives back $f$ and the second term is equal to $f$. Therefore, $T = \{e^x\cos x,e^x\sin x\}$ is the final answer, consistent with $\int e^x\sin x\,dx = (e^x\sin x - e^x\cos x)/2$. The main idea is repetitive differentiation of the terms until we run out of new terms to process. Essentially, we are generating a \emph{closure} of the input expression based on differentiation. 

Every $f \in \mathbb{C}[e^x,e^{-x}]$ has a finite number of candidate terms; therefore, the preceding algorithm always terminates. We prove this for functions defined according to Eq.~\ref{eq:def_exact}. After differentiation, each resulting term is also in the form of Eq.~\ref{eq:def_exact} and $p_1$ remains the same in all of them. In addition, $p_2 + p_3$ remain the same (one power of $\sin x$ converts to $\cos x$ or vice versa). Similarly, $p_4 + p_5$ is a constant. Therefore, the sum of the powers, $p = \sum_k p_k$, is a constant. Since for each $p$ there are only a finite number of possible expressions, we run out of new terms in a finite number of steps and the algorithm terminates. 

In general, most integrands are not in $\mathbb{C}[e^x,e^{-x}]$ and candidates cannot all be found by just repetitive differentiation; however, repetitive differentiation is still the backbone of the general algorithm. The essence of candidate generation is integration by parts,
\begin{equation}
    \int u v'\,dx = u v - \int u' v\,dx
    \,
    \label{eq:int_by_parts}
\end{equation}
followed by repetitive differentiation. This process or something equivalent resides at the core of most symbolic integration packages. However, when applying this algorithm to complex expressions, multiple problems arise. At each step, there can be more than one possible split of the integrand into $u$ and $v'$. It is unknown a priori which split will result in the correct solution. Therefore, we need to explore all possibilities. Doing this naively can generate an exponential number of candidates. In fact, there is no guarantee that the process even terminates. Realizing that many intermediate expressions are shared among different paths, we can manage the number of candidates by keeping track of the intermediate results, similar to how memoization reduces the complexity of recursive algorithms. A top-down recursive algorithm with memoization usually has an equivalent bottom-up dynamic programming dual. This is the algorithmic basis of the approach discussed below.

A more systematic way to manage the intermediate expressions generated by repetitive integration by parts is provided by the continuous homotopy operators, which automate integration by parts and are useful in integrating exact expressions composed of multiple functions of independent variables and finding conservation laws in the system of differential equations~\cite{Hereman2005, Deconinck2009, Poole2010}. Here, our goal is to borrow some of the machinery of the homotopy operators methodology to enhance the generation of candidate expressions. We continue with a brief and incomplete discussion of the homotopy operators technique. 

Let $\mathbf{u} = (u^1, u^2,\cdots,u^N)$ be a vector of $N$ dependent variables of $x$. Let $f(\mathbf{u})$ be a function of $\mathbf{u}$ and its partial derivatives with respect to $x$. We define $u_{0x}^i = u^i$, $u_x^i = \partial{u^i}/\partial{x}$ and $u^i_{kx} = \partial^k{u^i}/\partial{x}^k$. The total derivative operator is defined as
\begin{equation}
    \mathcal{D}_xf = \frac{\partial{f}}{\partial{x}} +
    \sum_{j=1}^{N} \sum_{k=0}^{M^j} u^j_{(k+1)x} \frac{\partial{f}}{\partial{u^j_{kx}}}
    \,
\end{equation}
where $M^j$ is the maximum order of the partial derivatives of $u^j$ present in $f$. The key to the homotopy operators technique is a method to invert $\mathcal{D}_xf$, namely $\int f\,dx = \mathcal{D}_x^{-1}f = \mathcal{H}_{\mathbf{u}}f$, where 
\begin{equation}
    \mathcal{H}_{\mathbf{u}}f = \int_{\lambda_0}^1
        \left(
            \sum_{j=1}^N \mathcal{I}_{u^j} f
        \right)
        [\lambda \mathbf{u}] \frac{d\lambda}{\lambda}
        \label{eq:Hu}
\end{equation}
is the one-dimensional homotopy operator, $\lambda$ is a dummy integration variable, and the notation $X[\lambda \mathbf{u}]$ means replacing $u^i$ with $\lambda u^i$, $u_x^i$ with $\lambda u_x^i$, and the same for other variables and partial derivatives in $X$. Moreover, $\mathcal{I}_{u^j}$ is defined as
\begin{equation}
    \mathcal{I}_{u^j}f = \sum_{k=1}^{M^j} 
        \left(
            \sum_{i=0}^{k-1} u^j_{ix} (-\mathcal{D}_x)^{k-(i+1)}
        \right)
        \frac{\partial{f}}{\partial{u^j_{kx}}}
        \,.
        \label{eq:Iu}
\end{equation}

For the proof of these equations, refer to \cite{Hereman2005, Poole2010}. However, we can make intuitive sense of these equations. Eq~\ref{eq:Iu} has two main roles. First, it replaces $u_{kx}$ (removed by the partial differentiation operator) with $u_{(k-1)x}$. This action is similar to integration by part, and, in fact, its proof is based on integration by parts. Second, it performs repetitive differentiation (by the sub-expression containing the powers of $\mathcal{D}_x$), reminiscent of the method discussed above for functions in $\mathbb{C}[e^x,e^{-x}]$. Finally, the definite integral in Eq~\ref{eq:Hu} on variable $\lambda$ calculates the coefficients of the candidates. 

In this paper, we use sparse regression to find the coefficients; therefore, we do not need Eq~\ref{eq:Hu}. Our focus is on Eq~\ref{eq:Iu}, but we cannot use it in its current form because the expected integrands are generally not in the form of $f(\mathbf{u})$. The key is to rewrite the integrand, $f(x)$, is a usable form,
\begin{equation}
    \boxed{
        f(x) = \prod_i^N u_i(x)^{n_i}\,,\, \text{where }\, u_i(x) = g_i(v_i(x))
        \,,
    }
    \label{eq:prod_u}
\end{equation}
$g_i(x)$ is a function that can be easily integrated (see Section 3.3), and $n_i \in \mathbb{Z}^+$. For example, if $f(x) = \sin^2 (x^2-1)$, then $f = u_1^2$ for $u_1 = \sin(v_1)$ and $v_1 = x^2-1$. 

To find candidates, we follow the same process as in Eq~\ref{eq:Iu} but customized for expressions conformant to Eq~\ref{eq:prod_u}. However, we cannot simply apply Eq~\ref{eq:Iu} because, for a general non-exact expression, $M^j$ may not be known or even bounded. Instead, we unroll the outer summation in Eq~\ref{eq:Iu} to generate an ordered list of candidate generators, $G_0, G_1, G_2,\cdots$. We generate $G_0$ by integrating each $u_i$ in turn. We integrate the first factor of $f$, i.e. $u_1^{n_1}$, by multiplying $f$ by $v_1'/v_1'$ and split $f$ into $u_1 v_1'$ and $f / (u_1 v_1')$. Ignoring the constant factors, the second part can be written as $(v_1')^{-1} \partial{f}/\partial{u_1}$. Considering that $u = g(v)$, we have $\int u v'\,dx = \int g(v) v'\,dx = \int g(v)\,dv$ (remember that $g$ is chosen to be easy to integrate). Therefore,
\begin{equation}
    \boxed{
        G_0 = 
        \sum_{i=1}^N
            \left(
                1 + 
                \int g_i(v)\,dv
            \right)_{v \gets v_i}
            \left(   
                1 + 
                v_i^{-1}
                \frac{\partial{f}}{\partial{u_i}}
            \right)
            \,.
    }
    \label{eq:B0}
\end{equation}
In $\left(1 + \int g_i(v)\,dv\right)$, 1 represents the constant of integration. For the term on the right, 1 is not strictly necessary but improves performance by moving terms from $G_{i> 0}$ to $G_0$. Next, we generate $G_1, G_2, \cdots$ by repetitive differentiation similar to Eq~\ref{eq:Iu}. In addition, we introduce the powers of $x$ into the results by integration-by-parts, assuming $v' = 1$ and $v = x$ in Eq~\ref{eq:int_by_parts},
\begin{equation}
    \int f\,dx = \int (x)' f\,dx = xf - \int x f'\,dx
    \,
    \label{eq:adding_x}
\end{equation}
Putting all together,
\begin{equation}
    \boxed{
        G_{i+1} =
            (1 + x)
            \left(
                \mathbf{1} +
                \mathcal{D}_x
            \right)
            G_i
            \,.
    }
    \label{eq:Bi}
\end{equation}

Let's apply the algorithm to $f(x) = \cot^4 x$. Decomposing $f$ based on Eq~\ref{eq:prod_u}, we have $u = \cot v$, $v = x$, and $f = u^4$. Considering that $\int \cot v\,dv = \log(\sin v)$, we get $G_0 = 1 + \log(\sin x) + \cot^3 x + \log(\sin x)\cot^3 x$, which does not solve the integral. Next, we move to $G_1$ by using Eq~\ref{eq:Bi}. This time, we have
\begin{dmath}
    G_1 = 1 + \log(\sin x) + \cot^3 x + \log(\sin x)\cot^3 x + x + \cot x +\cdots \text{(14 terms)}
    \,,
\end{dmath}
Feeding $f$ and the candidate set derived from $G_1$ to Algorithm~\ref{alg:num}, we solve the integral,
\begin{equation}
    \int \cot^4 x\,dx = x + \cot x - \frac{1}{3}\cot^3 x
    \,.
\end{equation}

\subsection{Rule-Based Integration}

Converting integrands to a form compatible with Eq.~\ref{eq:prod_u} and finding $\int g_i(v)\,dv$ in Eq.~\ref{eq:B0} are facilitated by the term-rewriting and rule definition functionality of \textbf{JuliaSymbolics}, which includes a layered architectural design \cite{cheli2021automated} to decouple term rewriting from other features, and leverages the Julia's metaprogramming features that have been inherited from other languages in the Lisp family. 

On the lowest level, \textbf{TermInterface.jl} and \textbf{Metatheory.jl} packages~\cite{Cheli2021, gowda2021high, cheli2021automated} provide an efficient expression rewriting engine and rule definition eDSL. Rules are written as regular Julia expressions, thus parsed into S-expressions and then efficiently compiled to callable functions that perform pattern matching and   support advanced conditional selection~\cite{ma2021modelingtoolkit, gowda2021high, cheli2021automated}.
\textbf{JuliaSymbolics} also supports the definition of equational (bidirectional) rewrite rules that can be executed through an e-graph rewriting engine (adopted from~\cite{10.1145/3434304}), performing equality saturation. 

We use \textbf{JuliaSymbolics}'s rewriting capabilities to find the integrals of $g(x)$. Nearly fifty rules are sufficient to cover elementary functions (exponential, logarithmic, trigonometric, hyperbolic, and their inverses). The following code snippets show a few example rules.

\begin{verbatim}
  @rule integrate(^(~x, -1)) => log(~x)
  @rule integrate(^(~x, ~k)) => ^(~x, ~k+1)
  @rule integrate(^(sin(~x), ~k::is_neg)) => 
         integrate(^(csc(~x), -~k))
  @rule integrate(csc(~x)) => 
         log(sin(~x)^-1 - cos(~x)*sin(~x)^-1)
\end{verbatim}

\section{Numerical Computation}

\subsection{Simplifying Rational Expressions}

Processing rational expressions is one of the first steps in symbolic integration with a long and mathematically rich history (e.g., Hermite's and Harowitz' methods). In this paper, to stay with the general theme of our method, we use a symbolic-numeric (hybrid) algorithm to factor rational expressions based on finding the roots of the denominator (which is assumed to be a polynomial). Our method is broadly similar to the one proposed in~\cite{NodaMatu-Tarow1992}; however, to find the coefficients of the terms, we use sparse regression method (see below) instead of the method of residues.

\subsection{Generating Random Points and Moving Toward the Poles}

In line 2 of Algorithm~\ref{alg:num}, $n$ random number $x_i$ (test points) are generated in, $\mathbb{D}_d$, an open disk of radius $d$ centered at the origin ($d$ is a parameter provided by the user). Therefore, $\mathbb{D}_d$ should contain the majority of the unique poles of $f(x)$. Note that a function like $\sec(x)$ has an infinity number of poles and $\mathbb{D}_d$ should contain at least one of them. For the sparse regression to work properly, test points need to be near the poles of $f(x)$.  We use a variant of the Newton-Raphson method to facilitate probing the poles. The standard Newton-Raphson method to find zeros of $f(x)$ is the iteration $x^{j+1} = x^j - f(x^j)/f'(x^j)$, where superscript depicts an order index and not a power. The zero-finding algorithm can be transformed to a pole-finding one by changing the minus sign to plus. Therefore, 
\begin{equation}
    x_i^{j+1} = x_i^j + \frac{f(x_i^j)}{f'(x_i^j)}
    \,,
    \label{eq:poles}
\end{equation}
is used by Algorithm~\ref{alg:num} on line 4 to move the test points toward the poles.

\subsection{Finding Linearly-Independent Subsets of the Candidates}

Matrix $\mathbf{A}$ in Algorithm~\ref{alg:num} can be rank-deficient. In order to find acceptable regression solutions, we need to remove linearly-dependent columns from $\mathbf{A}$. This task is achieved by using pivoted QR algorithm~\cite{Golub2013}. In short, $\mathbf{A}$ is decomposed to $\mathbf{A} = \mathbf{Q}\mathbf{R}\mathbf{P}^T$, where $\mathbf{Q}$ is an orthogonal rotation matrix, $\mathbf{R}$ is an upper triangular matrix, and $\mathbf{P}$ is the permutation matrix. We find the columns linearly dependent on other columns by locating small values on the diagonal of $\mathbf{R}$ (here, small means having an absolute value less than a parameter $\epsilon$, usually taken to be $10^{-6}$) and convert $\mathbf{P}$ to the set $R$, used by Algorithm~\ref{alg:num}, that lists indices of linearly-dependent columns. 

\subsection{Sparse Regression}

In Algorithm~\ref{alg:num}, we need to find $\mathbf{q}$ such that $\mathbf{A}\mathbf{q} = \mathbf{b}$. If $\mathbf{A}$ is low dimensional, this can be done simply as $\mathbf{q} = \mathbf{A}^{-1}\mathbf{b}$ (by construction, $\mathbf{A}$ is a square matrix). However, this process has the drawback of tending to use all the terms, even those with numerically small coefficients, which obscures the results and differs from the expected answer to integration problems. We need a parsimonious (sparse) model such that $\mathbf{q}$ has the minimum number of non-zero elements while still solves $\mathbf{A}\mathbf{q} = \mathbf{b}$ with an acceptable accuracy. We can achieve this by recasting the problem as an optimization problem to solve 
\begin{equation}
    \min_q \lVert \mathbf{A} \mathbf{q} - \mathbf{b} \rVert_2^2 + \lambda \lVert \mathbf{q} \rVert_i
    \,,
\end{equation}
for $\mathbf{q}$, where $i \in \lbrace 1, 2~\rbrace$, and $\lambda$ is a regularization parameter. In this paper, we use the sequential thresholded least-squares (STLSQ) algorithm, which is a component of Sparse Identification of Nonlinear Dynamics (SINDy) technique~\cite{Brunton2016} and uses $\ell_2$-norm. This method has been chosen due to robust behaviour within the scope of many problems related to SINDy. However, other sparse regression algorithms, e.g., the least absolute shrinkage and selection operation (LASSO) using $\ell_1$-norm~\cite{Tibshirani1996}, Elastic-Net or SR3 are also possible candidates. The sparse regression code is provided by \textbf{DataDrivenDiffEq.jl}~\cite{datadrivendiffeq}.

\section{Results}

\textbf{SymbolicNumericIntegration.jl} is available at \url{https://github.com/SciML/SymbolicNumericIntegration.jl}. We tested the performance of the software using a collection of test integrals provided by RUBI based on classic calculus textbooks~\cite{Rich2018} (Table~\ref{tab:axiom}). 

\begin{table}
\center{
  \caption{Performance of \textbf{SymbolicNumericIntegration.jl}}
  \label{tab:axiom}
  \begin{tabular}{ccccc}
    \toprule
    Source & Success & Failure & Total\\
    \midrule
    Apostle~\cite{Apostol1967} & 123 & 26 & 149\\
    Hearn & 191 & 38 & 229\\
    Stewart~\cite{stewart1991calculus} & 220 & 77 & 297\\
    Timofeev~\cite{Timofeev} & 136 & 126 & 262 \\
    \bottomrule
\end{tabular}
}
\end{table}

We showcase the strengths and weaknesses of the symbolic-numeric integration algorithm with two examples (both from Apostle~\cite{Apostol1967}). The algorithm can successfully solve the following integral,
\begin{equation}
    \int \frac{\log x}{x\sqrt{1 + \log x}}\,dx = 
    \frac{2}{3}x\log x\sqrt{1 + \log x} - 
    \frac{4}{3}\sqrt{1 + \log x}
    \,.
    \label{eq:success}
\end{equation}
The reason for the success is that we can solve Eq.~\ref{eq:success} after substitution $u = 1 + \log x$, which transforms the integral to easily solvable $\int (u-1) / \sqrt{u}\,du$. However, the integration algorithm does not explicitly apply the substitution. Instead, the way that the candidate generation algorithm (section 3.2) creates new candidates implicitly covers the substitution. On the other hand, the algorithm fails to solve the following simple integral,
\begin{equation}
    \int \frac{1}{1 + 2\cos x}\,dx = 
    \frac{\log(\tan x/2 + \sqrt{3}) - \log(\tan x/2 - \sqrt{3})}{\sqrt{3}}
\end{equation}
Here, there is no simple substitution that can solve the integral. The standard solution is obtained by expressing $\cos x$ in term of the tangent of half-angle and then solve the resulting rational expression. While it is easy to explicitly add this procedure to the algorithm, it is not an automatic consequence of the candidate generation algorithm based on homotopy operators. However, the algorithm can solve the following two variants of the problem, 
\begin{equation}
    \int \frac{\sin x}{1 + 2\cos x}\,dx = -\frac{1}{2}\log(1 + 2\cos x)
    \,,
\end{equation}
because now the substitution $u = 1 + 2\cos x$ works, and,
\begin{equation}
    \int \frac{1}{1 + \cos x}\,dx = \frac{\sin x}{1 + \cos x}
    \,,
\end{equation}
since $\sin x / (1 + \cos x) = \tan(x/2)$ and the algorithm can implicitly apply the tangent of half-angle trick. 
\section{Conclusions}

\textbf{SymbolicNumericIntegration.jl} provides a proof-of-concept that symbolic-numeric integration, based on a combination of symbolic candidates generation and numerical sparse regression, is able to solve many standard integration problems. Symbolic integration systems are tightly coupled to the CAS they are build on. \textbf{JuliaSymbolics} has strong simplification, rule-based rewriting, differentiation and code generation capabilities -- all useful for the scientific machine learning workflow it is designed to support. We add the much needed feature of symbolic integration by building on the same strengths of this CAS. On the other hand, polynomials with rational or integer coefficients ($\mathbb{Z}[x]$ and $\mathbb{Q}[x]$), which are the workhorse of most traditional CAS, currently lack robust algorithmic support. Therefore, \textbf{SymbolicNumericIntegration.jl} employs the strengths of \textbf{JuliaSymbolics}, as used in Eqs.~\ref{eq:prod_u}, \ref{eq:B0}, and \ref{eq:Bi}, and covers for its shortcomings by using numerical methods. We believe that the features of strong auto-differentiation and numerical routines combined with a relatively weak classic symbolic algebra are not unique to \textbf{SciML} ecosystem and are shared by many recent systems gearing toward machine learning and expect that the symbolic-numeric methodology developed in this paper can be applicable to these other systems. 


\section*{Acknowledgements}
This material is based upon work supported by the National Science Foundation under grant no. OAC-1835443, grant no. SII-2029670, grant no. ECCS-2029670, grant no. OAC-2103804, and grant no. PHY-2021825. We also gratefully acknowledge the U.S. Agency for International Development through Penn State for grant no. S002283-USAID. The information, data, or work presented herein was funded in part by the Advanced Research Projects Agency-Energy (ARPA-E), U.S. Department of Energy, under Award Number DE-AR0001211 and DE-AR0001222. We also gratefully acknowledge the U.S. Agency for International Development through Penn State for grant no. S002283-USAID. The views and opinions of authors expressed herein do not necessarily state or reflect those of the United States Government or any agency thereof. This material was supported by The Research Council of Norway and Equinor ASA through Research Council project "308817 - Digital wells for optimal production and drainage". Research was sponsored by the United States Air Force Research Laboratory and the United States Air Force Artificial Intelligence Accelerator and was accomplished under Cooperative Agreement Number FA8750-19-2-1000. The views and conclusions contained in this document are those of the authors and should not be interpreted as representing the official policies, either expressed or implied, of the United States Air Force or the U.S. Government. The U.S. Government is authorized to reproduce and distribute reprints for Government purposes notwithstanding any copyright notation herein.


\bibliographystyle{unsrt}  
\bibliography{SymNum}

\end{document}